\documentclass[prl,aps,twocolumn,reprint,floatfix,superscriptaddress,longbibliography,amssymb]{revtex4-1}
\usepackage[utf8]{inputenc} 
\usepackage{graphicx}
\usepackage{amsmath}
\usepackage{float}
\usepackage{xcolor}
\newcommand{\be}{\begin{equation}}
\newcommand{\ee}{\end{equation}}
\newcommand{\bea}{\begin{eqnarray}}
\newcommand{\eea}{\end{eqnarray}}
\newcommand{\ben}{\begin{equation*}}
\newcommand{\een}{\end{equation*}}
\newcommand{\ba}{\begin{align}}
\newcommand{\ea}{\end{align}}

\usepackage{braket}
\usepackage{times}
\usepackage{kbordermatrix}

\begin{document}

\title{Creating quantum many-body scars through topological pumping of a 1D dipolar gas}

\author{Wil Kao}
\altaffiliation[W.K.~and K.-Y.~Li~contributed equally to this work.]{}
\affiliation{Department of Applied Physics, Stanford University, Stanford, CA 94305, USA}
\affiliation{E.~L.~Ginzton Laboratory, Stanford University, Stanford, CA 94305, USA}
\author{Kuan-Yu Li}
\altaffiliation[W.K.~and K.-Y.~Li~contributed equally to this work.]{}
\affiliation{Department of Applied Physics, Stanford University, Stanford, CA 94305, USA}
\affiliation{E.~L.~Ginzton Laboratory, Stanford University, Stanford, CA 94305, USA}
\author{Kuan-Yu Lin}
\affiliation{Department of Physics, Stanford University, Stanford, CA 94305, USA}
\affiliation{E.~L.~Ginzton Laboratory, Stanford University, Stanford, CA 94305, USA}
\author{Sarang Gopalakrishnan}
\affiliation{Department of Engineering Science and Physics, CUNY College of Staten Island, Staten Island, NY 10314, USA}
\author{Benjamin L.~Lev}
\affiliation{Department of Applied Physics, Stanford University, Stanford, CA 94305, USA}
\affiliation{Department of Physics, Stanford University, Stanford, CA 94305, USA}
\affiliation{E.~L.~Ginzton Laboratory, Stanford University, Stanford, CA 94305, USA}

\date{\today}

\begin{abstract}

Quantum many-body scars, long-lived excited states of correlated quantum chaotic systems that evade thermalization, are of great fundamental and technological interest.  We create novel scar states in a bosonic 1D quantum gas of dysprosium by stabilizing a super-Tonks-Girardeau gas against collapse and thermalization with repulsive long-range dipolar interactions. Stiffness and energy density measurements show that the system is dynamically stable regardless of contact interaction strength.  This enables us to cycle contact interactions from weakly to strongly repulsive, then strongly attractive, and finally weakly attractive.  We show that this cycle is an energy-space topological pump (due to a quantum holonomy). Iterating this cycle offers an unexplored topological pumping method to create a hierarchy of quantum many-body scar states.

\end{abstract}

\maketitle

Highly excited eigenstates of interacting quantum systems are generically ``thermal,'' in the sense that they obey the eigenstate thermalization hypothesis~\cite{rigol_review}: physical observables behave in these excited states as they would in thermal equilibrium. For generic thermal systems, all initial conditions give rise to locally thermal behavior at times past the intrinsic dynamical timescale. Systems in which thermalization is absent are of great fundamental interest, since they violate equilibrium statistical mechanics, and of technological interest since  some quantum information in these states evade decoherence. Nonthermal excited states exist in integrable~\cite{rdo} and many-body localized~\cite{abanin_rmp} systems; more recently, it has been realized that even nonintegrable systems might have special initial states for which thermalization is slow or absent. These states are called quantum many-body scars~\cite{Bernien:2017bp,sm2017, Turner:2018iz,Moudgalya:2018gt,Khemani:2019dw}; they are many-body analogs of certain eigenstates in chaotic billiards that ``remember'' their proximity to an unstable periodic orbit~\cite{Heller:1984bs}. So far, quantum many-body scars have been experimentally studied for only one fine-tuned initial state, in a specific lattice model with Rydberg atoms~\cite{Bernien:2017bp}; in that experiment, scars were manifest as long-lived magnetic oscillations. Theoretically, however, scar states (and hierarchies thereof) have now been found in a number of Hamiltonians~\cite{sm2017, Turner:2018iz,Moudgalya:2018gt,Khemani:2019dw}; these scars do not always manifest as oscillations, but more generally as observables that remain far from their thermal value for long times. Much about their physics remains unclear, and novel forms are of great interest including those presented here, which are the first  observed in a continuous, rather than lattice-based system. 

\begin{figure*}[t!]
\includegraphics[width = \textwidth]{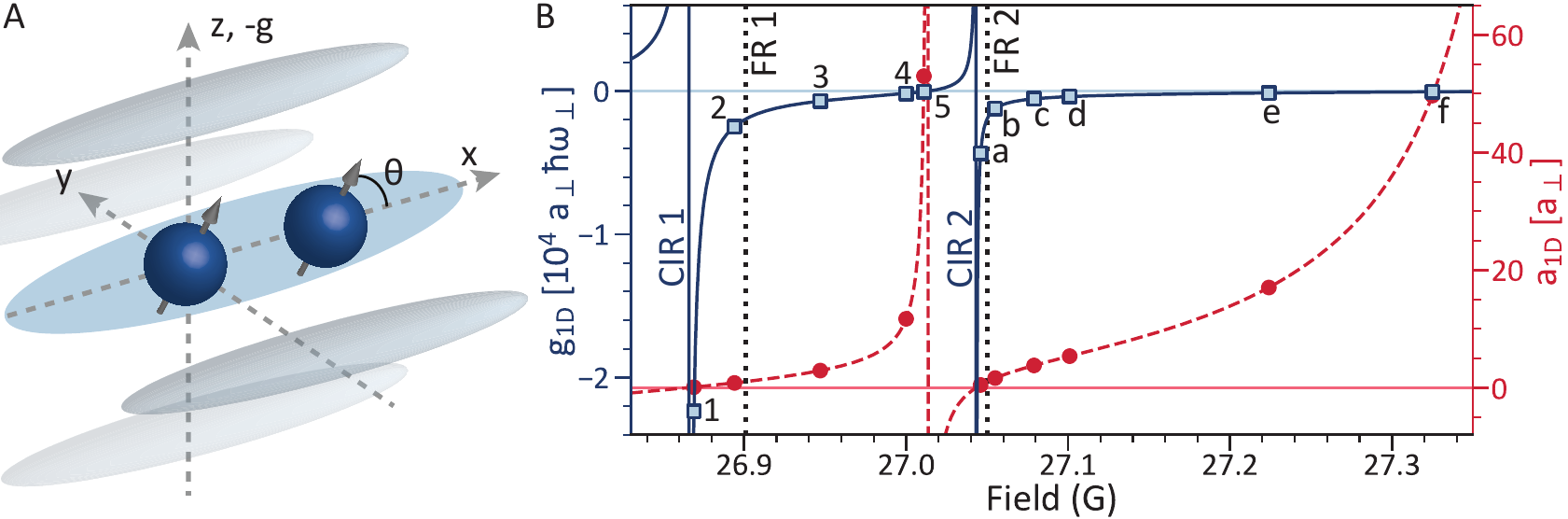}
\caption{(A) 1D traps formed by a 2D optical lattice.  Atomic dipoles are aligned by a magnetic field at angle $\theta$ from $\hat{x}$ in $x$--$z$ plane.  (B) The applied field  tunes the contact interaction strength $g_\text{1D}$ (solid) and 1D scattering length $a_\text{1D}$ (dashed) via two confinement-induced resonances (CIRs) located to the low-field side of Feshbach resonances (FRs) indicated by dotted lines.  For reference, $g_\text{1D}<0$ measurements are labeled by numbers and letters for the first and second holonomy cycles, respectively. }
\label{cartoons}
\end{figure*}

In this work, we demonstrate a ``topological'' pumping protocol for creating a hierarchy of quantum many-body scars, by cyclically varying the interaction strength of a dipolar Bose gas confined in one dimension. The cycles are made possible through dipolar stabilization of the gas.  In a conventional topological pump~\cite{thouless_pumping}, the Hamiltonian returns to itself after one cycle, but the state is translated by one lattice site. In the present setup, by contrast, the state is translated up the many-body energy spectrum; thus, each eigenstate is pumped to an eigenstate with an extensively higher energy. This phenomenon, which is a consequence of integrability, is called a ``quantum holonomy''~\cite{yonezawa}. In an intermediate stage of this cycle, the system forms an attractively interacting, metastable ``super-Tonks-Girardeau'' gas (sTG), in which the bosons are even more strongly anticorrelated than free fermions~\cite{Astrakharchik:2004ht,Astrakharchik:2005fz,Batchelor:2005gt,Haller:2009jrb,Chen:2010dl,Solano:2019ig}. As one quenches deeper into the sTG regime by making the interactions less strongly attractive, one expects the sTG to become unstable; this is indeed seen in gases with purely short-range interactions~\cite{Haller:2009jrb}. Remarkably, however, even though the dipole-dipole interaction (DDI) breaks integrability~\cite{Tang:2018dq}, it \emph{enhances} the stability of the sTG regime (relative to the purely short-range case). This allows one to implement the entire cycle, thus realizing this previously unobserved topological pumping phenomenon. The sTG regime at intermediate interaction strengths is a scar state. That such a route from approximate integrability to scars might exist was first pointed out in Ref.~\cite{Khemani:2019dw}. (Dipolar sTGs have been predicted to exist in contexts different from that realized here; see supplementary materials for discussion~\cite{Supp}.) 

We implement the following protocol. First, we create a low-temperature dipolar Bose gas in a regime with weak repulsive short-range (contact) interactions. We then tune the scattering length across confinement-induced resonances (CIRs) of colliding atoms~\cite{Olshanii:1998jr,Bergeman:2003kn,Moritz:2005bf,Haller:2010dj} in the following stages. First, we ramp-up the contact interactions toward the resonance, so the gas adiabatically enters the strongly antibunched Tonks-Girardeau (TG) state%, in which the bosonic wavefunction can be mapped onto that of free fermions
~\cite{Girardeau:1960ff,Paredes:2004fp,Kinoshita:2004jp}. At this point, we quench these interactions across the resonance, from strongly repulsive to strongly attractive to create the sTG. As the attractive interactions are tuned away from this unitary contact regime, the sTG gas usually becomes thermodynamically unstable because the bosons can form soliton-like bound cluster states~\cite{McGuire:1964dw,Astrakharchik:2004ht,Astrakharchik:2005fz}, as has been observed in a nondipolar gas~\cite{Haller:2009jrb}. Nevertheless, the dipolar system appears dynamically stable for very long times. We then ramp the attractive contact interaction strength toward zero again to generate a weakly attractive Bose gas in a highly excited nonthermal state. That the system remains dynamically stable throughout this procedure is a consequence of the repulsive dipolar interactions, as we will discuss below. Repeating the cycle by crossing another CIR produces even higher excited nonthermal scar states. These claims are supported through gas stiffness and energy density measurements at various stages in the protocol.

We begin our experiments by loading a nearly pure Bose-Einstein condensate (BEC) of highly magnetic Dy atoms into a 2D optical lattice~\cite{Tang:2018dq} whose  first transverse excited state energy is $\hbar\omega_\perp/k_\text{B}=1180(20)$~nK. ($^{162}$Dy's magnetic moment of $\mu=10$ Bohr magnetons is 10$\times$ that of, e.g., Cs's, yielding a DDI ${\sim}$100$\times$ stronger.)  This forms an array of ${\sim}1000$ 1D optical traps with about 30 atoms in the central tube and 20 atoms per tube on average; see Fig.~\ref{cartoons}A and Ref.~\cite{Supp}. Each tube approximates a 1D channel of finite length: the ratio of  longitudinal versus  transverse oscillator lengths is $a_\parallel/a_\perp = 25$. The transverse trap frequency is $\omega_\perp=2\pi\times 24.6(4)$~kHz and $h=2\pi\hbar$ is Plank's constant. Collective oscillation measurements are consistent with zero-temperature ground state predictions~\cite{Supp}, which implies that the temperature is sufficiently low to observe the sTG gas~\cite{Kormos:2011eb}.
 
The system may be described with a Lieb-Linger (LL) Hamiltonian~\cite{giamarchi2003quantum,Cazalilla:2011dm} augmented by the magnetic DDI:
\bea
H=-\frac{\hbar^2}{2m}\sum^N_{j=1}\frac{\partial^2}{\partial x_j^2}+g_\text{1D}\sum_{1\leq i < j \leq N}\delta(x_i{-}x_j)+ \nonumber V^\text{1D}_\text{DDI}(\theta,x),
\eea
where the first two terms comprise the LL model and the third is the 1D-regularized DDI~\cite{Supp}. Because the DDI scales as $1-3\cos^2{\theta}$, we can control its sign and  strength  by applying an external magnetic field $\mathbf{B}$ to polarize the dipoles at an angle $\theta$ with respect to the 1D axis $\hat{x}$.   The contact interaction strength $g_\text{1D}$ is independently controlled by setting the field magnitude $B$ to be near a CIR while holding $\theta$ constant; see Fig.~\ref{cartoons}B.

A CIR appears when the bound state of the first transverse motional excited state of the 1D trap is degenerate with the open-channel transverse ground state. They modify the contact interaction strength by
\be\label{CIRtheory}
g_\text{1D}(B) = -\frac{2\hbar^2}{ma_\text{1D}(B)}=\frac{2\hbar^2 a_\text{3D}(B)}{ma_\perp^2}\frac{1}{1-C a_\text{3D}(B)/a_\perp}.
\ee
Here, $C \approx 1$ and $a_\text{3D}$ and $a_\text{1D}$ are the 3D and 1D scattering lengths, respectively~\cite{Bergeman:2003kn}.  We tune $g_\text{1D}$ by controlling $a_\text{3D}$ with a Feshbach resonance (FR) at fixed $a_\perp$.  Feshbach resonances provide a means for tuning $a_\text{3D}$ via control of $B$~\cite{Chin:2010kl}.  We prepare the BEC at a field 200-mG below two Feshbach resonances around 27~G~\cite{Lucioni:2018ir,Supp} and find CIRs to the low-field side of each; see Fig.~\ref{cartoons}B. The gas enters the unitary contact-interaction regime $\gamma\rightarrow \pm \infty$ when $B$ sets $a_\text{3D} = a_\perp/C$~\cite{DeRosi:2017dj}.  The dimensionless LL parameter is $\gamma = g_\text{1D}m/\hbar^2n_\text{1D}$, with $n_\text{1D}$ the 1D atomic density.
 
While the DDI has been predicted to affect CIRs~\cite{Sinha:2007gx,Supp}, we resolve no shift of these resonances' positions or widths versus $\theta$ in our molecular bound state measurements~\cite{Supp}. Indeed, their position is adequately predicted by the nondipolar theory result of Eq.~\eqref{CIRtheory}. This simplifies the mapping of $B$ to $a_\text{3D}$ (and hence to $g_\text{1D}$) by rendering it $\theta$-independent.  We  prepare states with a particular $g_\text{1D}$ and implement the holonomy cycle(s)  by sweeping $B$ up to the desired higher-field value. The second holonomy cycle begins after point 5 in Fig.~\ref{cartoons}, where $g_\text{1D}$ turns positive again, and continues to point \textit{f}.

\begin{figure}[t!]
\includegraphics[width = \columnwidth]{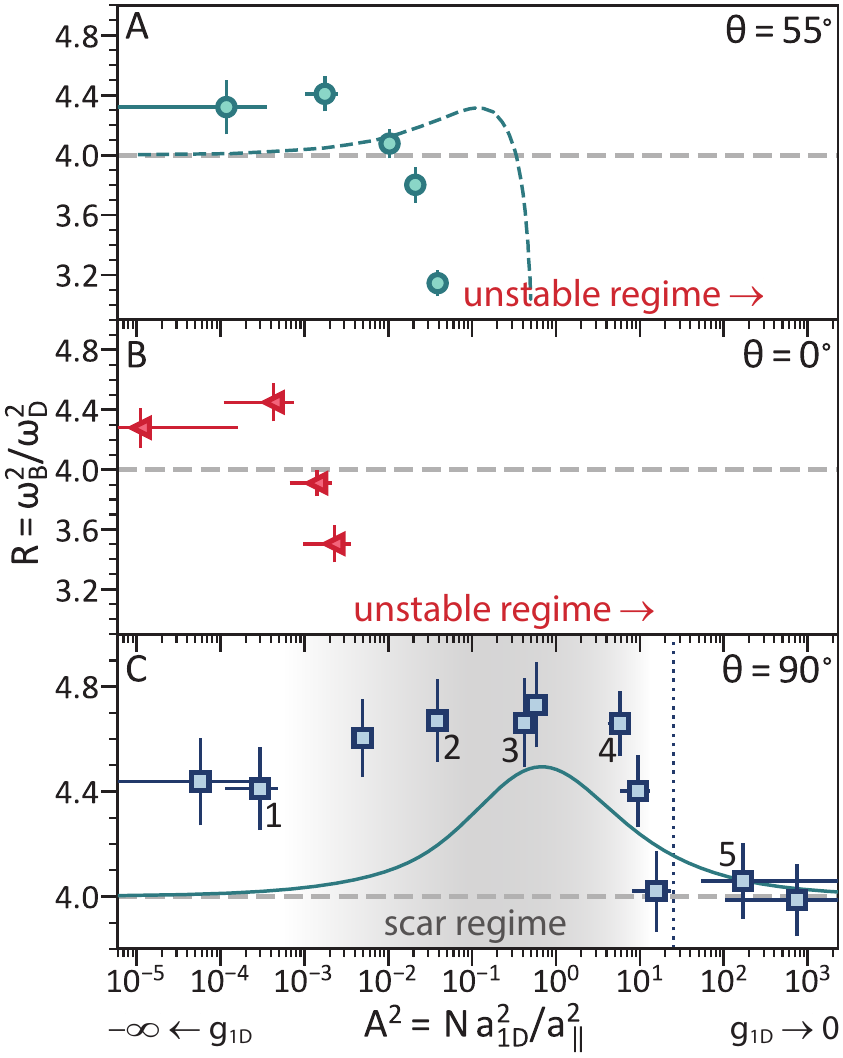}
\caption{Post-quench gas stiffness $R$ versus interaction parameter $A^2$ in the attractive  $g_\text{1D}<0$ regime of the first holonomy cycle. Measurements are shown for the nondipolar ($\theta=55^\circ$) and attractive DDI ($0^\circ$) systems in panels A and B, respectively, and for the repulsive, $90^\circ$ DDI-stabilized excited gas in panel C.  In (A) and (B), an sTG gas exists in the unitary regime of $A^2 \alt 10^{-3}$.  Beyond, however, the gas softens before collapsing near $A^2 \approx 10^{-1}$ and $10^{-2}$, respectively. For comparison, the dashed green curve in panel A plots data from the nondipolar variational Monte Carlo simulation of Ref.~\cite{Astrakharchik:2005fz}. (C) Surprisingly, the repulsive DDI system remains stable beyond the unitary regime.  This allows scar states to emerge around intermediate coupling strengths, indicated in gray, before crossing over into the $R=4$ weakly attractive, excited Bose gas regime beyond $A^2 \approx 10$. The solid curve is the Bethe ansatz prediction. The vertical dotted line indicates where the contact and the short-range 1D-regularized DDI contributions become approximately  equal~\cite{Supp}.  Numbers refer to points in Figs.~\ref{cartoons}B and~\ref{energy}. The error bars here and in subsequent figures represent the standard error.}
\label{stiffness}
\end{figure}

We measure gas stiffness via observations of collective oscillations of the atoms along the 1D trap axis~\cite{Menotti:2002jc,Astrakharchik:2005fq,Haller:2009jrb}. The frequency $\omega_\text{B}$ of the breathing mode of the gas width is sensitive to its inverse compressibility (stiffness). Normalizing $\omega_\text{B}$ by the frequency $\omega_\text{D}$ of the center-of-mass dipole (sloshing) mode accounts for nonuniversal aspects of the 1D potentials, such as trap frequencies~\cite{Astrakharchik:2005fq}.  This allows one to compare the stiffness of disparate systems at different interaction strengths by plotting $R= (\omega_\text{B}/\omega_\text{D})^2$ versus $A^2 = Na_\text{1D}^2/a_\parallel^2$. $A$ is the universal form of the coupling constant  under the local density approximation~\cite{Astrakharchik:2005fq}. At strong coupling ($g_\text{1D} \rightarrow \pm \infty$), $A^2\rightarrow 0$, while $A^2$ diverges at weak coupling ($g_\text{1D} \rightarrow 0^\pm$).  To measure oscillations, we selectively excite one of these two modes and hold the gas for a variable amount of time before releasing to image its width or center-of-mass in  time-of-flight. We repeat for the other mode and fit the oscillations to extract collective mode frequencies and $R$ values~\cite{Supp}. 

Figure~\ref{stiffness} shows stiffness data for excited states of the attractive ($g_\text{1D}<0$) dipolar LL model at three different $\theta$.  (Data for ground states of the repulsive $g_\text{1D}>0$ model are in Ref.~\cite{Supp}.)  We begin with the nondipolar case of $\theta = 55^\circ$ (at which the DDI vanishes along the 1D tube) in Fig.~\ref{stiffness}A, so as to compare to prior nondipolar Cs measurements~\cite{Haller:2009jrb} and to nondipolar theory~\cite{Astrakharchik:2005fz}. After preparing a TG gas (at which $R=4$) by tuning $\gamma\rightarrow +\infty$~\cite{Supp}, we quench into the attractive contact regime where $\gamma\rightarrow -\infty$, $A^2 \rightarrow 0$, and the TG gas crosses over into the sTG gas. Tuning $A^2$ larger causes the stiffness to rise above $R=4$, indicating that a stiffer (more strongly correlated) sTG gas forms; see panel A. At still larger $A^2$, $R$ rapidly decreases as the gas softens, indicating an imminent collapse into bound cluster states. This trend resembles that reported for the Cs system~\cite{Haller:2009jrb,Supp}, though we additionally report metastable states just below $R=4$: These might be gas-like states of  clusters of two or three bound atoms~\cite{Batchelor:2005gt}.  

Why should this nondipolar gas collapse, given that the attractive LL model remains integrable for all $A^2$? In the strictly integrable limit, collapse does not occur, and instead the stiffness rises above $R=4$ till $A^2 \approx 1$, then decreases to 4 in the weakly attractive regime~\cite{Chen:2010dl}. Many-body states with and without clusters belong to separate sectors of Hilbert space, and do not mix. In realistic experiments (including nondipolar ones~\cite{Haller:2009jrb}), imperfections such as the transverse and longitudinal trap potentials break integrability~\cite{Mazets:2010hk,Tan:2010ha} and yield matrix elements (proportional to the wavefunction overlap) between the sTG state and the bound cluster states, leading to collapse. In the strongly interacting unitary limit $A^2 \alt 10^{-3}$, antibunching strongly suppresses wavefunction overlap, so the model remains nearly integrable and stable despite experimental imperfections. However, in the intermediate interaction regime, cluster states form and the nondipolar gas is dynamically unstable, as seen in Fig.~2A.

\begin{figure}[t!]
\includegraphics[width = \columnwidth]{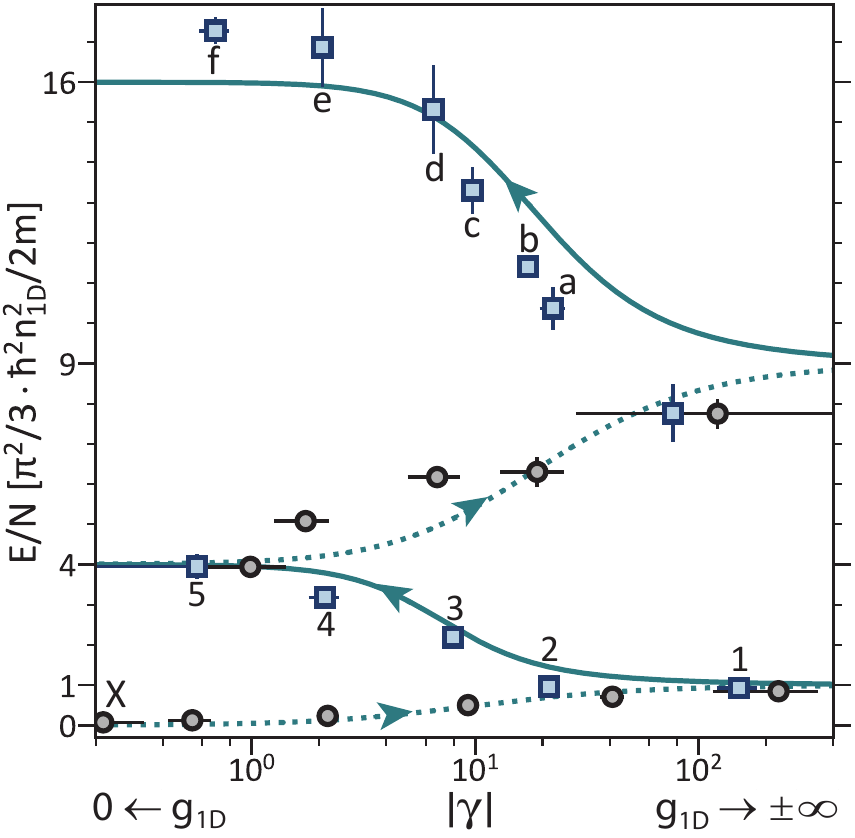}
\caption{Energy eigenstate spectrum plotted as energy per particle $E/N$ of states topologically pumped up two complete quantum holonomy levels. Black circles (blue squares) are data taken for positive (negative) $g_\text{1D}$'s of the repulsive (attractive) LL model.  Likewise, dotted (solid) curves are solutions to the Bethe ansatz equations for the repulsive (attractive) LL model. In the $g_\text{1D}\rightarrow 0$ and $\pm\infty$ limits, these solutions equal integer multiples of $1/3$ times the Fermi energy $\hbar^2(\pi n_\text{1D})^2/2m$. Arrowheads indicate direction of cycles.  The first cycle begins at the point labeled `\textit{X}' and continues to  point 5, where the second cycle begins and continues to \textit{f}.  The system passes through $|\gamma|\rightarrow\infty$ twice as the field increases first through CIR 1, then CIR 2. }
\label{energy}
\end{figure}

\begin{figure}[t!]
\includegraphics[width = \columnwidth]{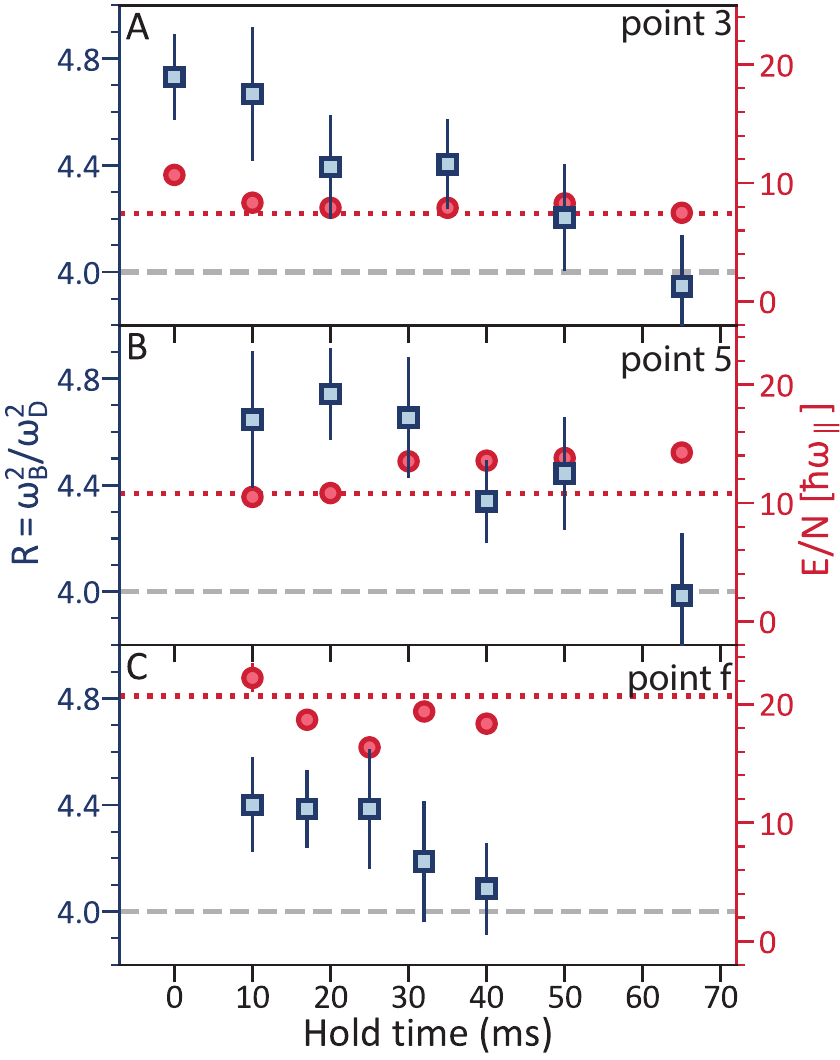}
\caption{Stiffness $R$ (squares) and energy per particle $E/N$ (circles) versus hold time for excited-state gases prepared in the attractive first and second holonomy levels. (A) Data for the scar state at point 3.   (B) and (C) show similar data for weakly attractive excited states at points 5 and \textit{f}, respectively.  
The energy stays approximately constant at the Bethe ansatz prediction (red dotted line) in all cases.  This implies that the decay in stiffness toward the thermal-state value $R=4$ is due purely to thermalization and not to heating (i.e., to an increase in entropy rather than energy). Thus, atypical excited states can be prepared and persist for long times in both the scar and weakly interacting regimes. } 
\label{thermalization}
\end{figure}

The collapse ensues even earlier if an attractive DDI is introduced by rotating to $\theta = 0^\circ$:  Figure~\ref{stiffness}B  shows a collapse beginning at roughly an order-of-magnitude lower in $A^2$. Evidently, the attractive DDI acts to break integrability at a point deeper within the unitary regime. Indeed, previous work showed that DDIs  generically break integrability~\cite{Tang:2018dq}.  From this perspective, one might also expect an early collapse under a repulsive DDI. Surprisingly, however, this is not the case: The  $\theta=90^\circ$ data in Fig.~\ref{stiffness}C  shows that the sTG remains stable orders-of-magnitude beyond that of the nondipolar gas. In fact, the gas \textit{never} collapses:  $R\agt4$ no matter how weak the interactions become at large $A^2$.  This is true even as the contact becomes weaker than the DDI,  vanishes, and become positive again as CIR 2 is approached within the second holonomy cycle.  (By contrast, dipolar BECs in higher dimensions collapse whenever the attractive contact exceeds the DDI~\cite{Lahaye2009}.)

How the repulsive DDI inhibits the sTG eigenstate from mixing with cluster eigenstates is unclear, though variational Monte Carlo provides intuition.  Simulations of a nondipolar gas in a harmonic trap of finite $a_\perp$ exhibit an energy barrier to collapse that shrinks as $A^2$ grows~\cite{Astrakharchik:2004ht}. Intuitively, then, the repulsive (attractive) DDI may serve to raise (lower) this barrier.  This is despite the relatively low DDI energy scale, which does not seem to change the TF-to-TG crossover of the ground states of the repulsive LL model~\cite{Supp}. Nevertheless, the DDI \textit{does} have a dramatic affect on the stability of the \textit{excited} states, rendering their $R$-dependence quite similar to that predicted by the Bethe ansatz equations~\cite{Chen:2010dl}.  Quantitative agreement is not expected as the Bethe ansatz solutions exclude effects due to, e.g., the DDI, trap, and imperfect state preparation.

Next, we explore those states obtained by sweeping $B$ past the second CIR, thereby entering the second holonomy cycle in $g_\text{1D}$ as reflected in the eigenenergy spectrum~\cite{yonezawa}.  Energy per particle $E/N$ is measured in time-of-fight by absorption imaging. The average momentum is determined from the expansion time and gas width, ensemble averaged over the 1D trap array~\cite{Supp}. This is shown in Fig.~\ref{energy}, where $E/N$ is plotted versus $\gamma$ along with the eigenenergy bands derived from the Bethe ansatz equations~\cite{Chen:2010dl,yonezawa,Supp}. We see that crossing the second CIR maps the system to higher energy eigenstates than those at the same $g_\text{1D}$ in the previous topological pumping cycle.  The orientation of the quench cycle is crucial:  Reversing the sense of the cycle does not implement the topological pump, but leads to collapse~\cite{yonezawa,Supp}.  The measured energies are in surprisingly good agreement with the Bethe ansatz predictions.

We identify the new excited states lying between the unitary sTG  and weakly interacting regimes as forming a hierarchy of quantum many-body scars; i.e., those within approximately $10^{-3} < A^2 <10$ ($1< \gamma < 10^2$). These are carefully prepared, long-lived, highly excited many-body states in a nonintegrable system that exhibit persistent athermal behavior. To establish this, we now demonstrate they are atypical and distinct from thermal states at the same energy density. We do so by observing both $R$ and the energy of these states as a function of time held in that state. We begin with point 3. Figure~\ref{thermalization}A shows that while $R$ slowly decreases to the $R=4$ value of a thermal state~\cite{Moritz:2003kv}, its $E/N$ stays the same. Thus, in becoming a thermal state, the gas must be increasing its entropy because it does not heat. This implies that the initial strongly correlated state with $R=4.7$  is atypical, yet sufficiently long-lived to be well-characterized by collective oscillation measurements. 

To show that the gas is nonthermal throughout the holonomy cycles, we repeat this measurement for two weakly interacting excited states: points 5 and  \textit{f} in the first and second attractive holonomy levels; see Figs.~\ref{thermalization}B and C, respectively.  Bethe ansatz predicts these states will have $R$'s close to 4 (confirmed in Ref.~\cite{Supp}), unfortunately just like thermal states.  So to distinguish them from thermal states using stiffness, we wait to measure $R$ not at their $g_\text{1D}$, but after we have swept $B$ back to a state (point 3) whose stiffness is naturally above 4.  Such $R$ measurements should remain greater than 4 as long as the excited state held at point 5 or \textit{f} is nonthermal. As shown in Figs.~\ref{thermalization}B and C, we indeed find that neither gas immediately relaxes to an $R=4$ thermal state nor significantly heats. Being weakly interacting, however, we do not consider these states scars.

We experimentally realized a spectral topological pump that takes a Bose gas from its ground state to a ``ladder'' of nonthermal excited scar states; we demonstrated the first two steps on this ladder. The central topological feature of this pump---quantum holonomy---is only strictly well-defined in the integrable limit~\cite{Khemani:2019dw}. The imperfections in any realistic experiment imply that one is never at the integrable limit, and render the system unstable. Remarkably, however, dipolar interactions (which themselves break integrability) stabilize the sTG against the effects of other integrability-breaking perturbations, and increase the lifetime of the scar states long enough that one can complete a full cycle and thus experimentally utilize the holonomy. The observation that two integrability-breaking perturbations can counteract each other might have wide-ranging implications for understanding the onset of chaos in nearly integrable quantum systems~\cite{begr,Tang:2018dq,mrdr,fgv}. Future work can  characterize these excited states through measurements of dynamic susceptibility and lifetime versus $\theta$. Moreover, because Feynman's `no-node' theorem does not apply to the wavefunctions of \textit{excited} bosonic states~\cite{Wu:2011bj}, these bosonic scars may realize exotic fermionic many-body systems; e.g., cotrapping Bose or Fermi impurity atoms may yield unusual polarons.

We thank Vedika Khemani, Gregory Astrakharchik, Eugene Demler, David Weiss, Cheng Chin, John Bohn, Dmitry Petrov, Maria-Luisa Chiofalo, Roberta Citro, and Stephanie De Palo for stimulating discussions.  We acknowledge funding support from the NSF and AFSOR.  W.K.~and K.-Y.~Lin~acknowledge partial support from NSERC and the Olympiad Scholarship from the Taiwan Ministry of Education, respectively. 

%\bibliographystyle{apsrev4-1-prx}
%\bibliography{STG,LevLab,supp}

%merlin.mbs apsrev4-1.bst 2010-07-25 4.21a (PWD, AO, DPC) hacked
%Control: key (0)
%Control: author (72) initials jnrlst
%Control: editor formatted (1) identically to author
%Control: production of article title (-1) disabled
%Control: page (0) single
%Control: year (1) truncated
%Control: production of eprint (0) enabled
%

\setcounter{figure}{0}  

\renewcommand{\figurename}{Supp.~Fig.}

\onecolumngrid
\subsection*{{\Large{Supplementary Materials:}}\\\vspace{2mm} \large{Creating quantum many-body scars through topological pumping of a 1D dipolar gas}\\\vspace{2mm} \normalsize{\normalfont{Wil Kao*, Kuan-Yu Li*, Kuan-Yu Lin, Sarang Gopalakrishnan, and Benjamin L.~Lev}}}

\vspace{-3mm}
\begin{figure}[b!]
    \centering
    \includegraphics[width=\textwidth]{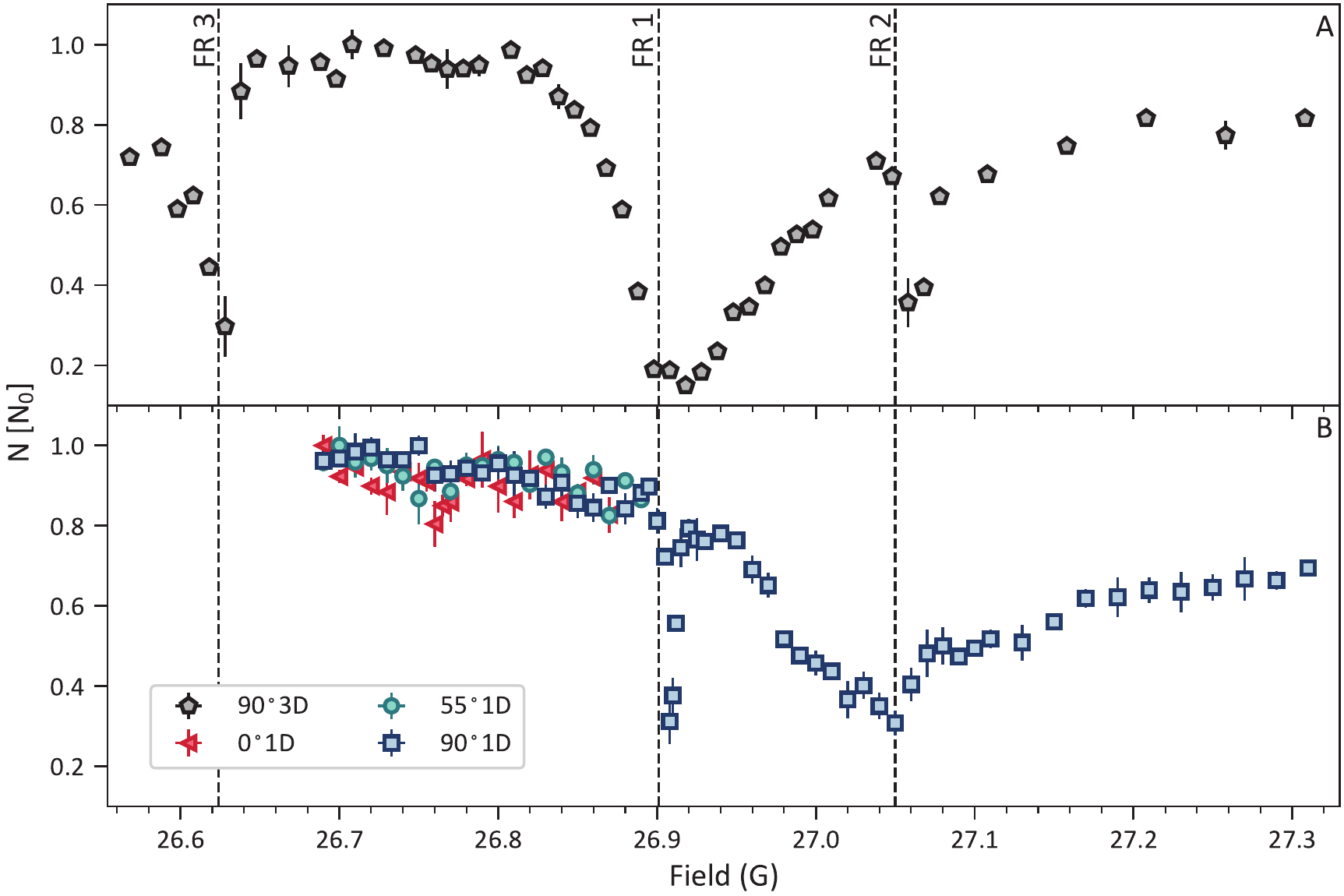}\vspace{-3mm}
    \caption{(A) Poles of the Feshbach resonances (FRs), indicated in dashed lines and determined from the molecular binding energy measurements, are overlaid on top of a high-resolution atom-loss spectrum taken using an ultracold thermal gas for a hold time of 50~ms in the lattice. (B) Atom-loss spectra in the 1D degenerate regime for the same lattice hold time are presented for the $B$-fields studied in Fig.~2 and Supp.~Fig.~\ref{fig:second_attractive} at each $\theta$. In both panels, the atom number $N$ is normalized by the peak number $N_0$ obtained in a field region away from any Feshbach resonance.}
    \label{fig:fr_loss}
\end{figure}

\vspace{2mm}
\section{Materials and methods} \vspace{-2mm}
\subsection{BEC production} \vspace{-2mm}
We follow the procedure in Ref.~\cite{Tang2015} to produce a $^{162}$Dy BEC in the Zeeman sublevel $m_J = -8$~$(J = 8)$, the absolute ground state. The atoms are loaded into a far-off-resonance single-beam optical dipole trap (ODT1) from a $741$-nm magneto-optical trap (MOT).  Instead of evaporating at $1.58$~G as in  Ref.~\cite{Tang2015}, we ramp the field to $26.69$~G in less than $1$~ms.  The scattering length at this field value is $150(6)\,a_0$, and we experimentally identified this field to be optimal for BEC production between the broad $22$-G and $27$-G Feshbach resonances; see Ref.~\cite{Lucioni2018} and Supp.~Fig.~\ref{fig:fr_loss}. The ramp sequence avoids having to sweep through the dense Feshbach spectrum of Dy with a condensed gas, where heating due to inelastic three-body collisions becomes significant. This protocol yields a nearly pure BEC of $2.23(5)\times 10^4$ atoms after the atoms are transferred from ODT1 into a crossed dipole trap (ODT2) for forced evaporation. At the end of evaporation, the ODT2 trap frequency is set to $[\omega_x,\omega_y,\omega_z] = 2\pi\times [63.9(6),11.8(3),166.4(5)]$~Hz. The magnetic field axis is kept along the direction of gravity $\hat{z}$ in the above procedure. We then rotate the field to align the dipoles at the desired $\theta$, where $\theta$ is the angle subtended by the dipole polarization and $\hat{x}$. The rotation is adiabatic such that no collective mode is excited. The amplitude of the bias field is kept constant so that it does not coincide with Feshbach resonance features during the rotation. This procedure minimizes three-body heating and atom loss.

\subsection{Quasi-1D confinement} 
Quasi-1D dipolar gases have been created in  1D tubes of 2D optical latices in systems using Cr in the weakly repulsive interacting regime~\cite{Pasquiou:2011dg} and Dy in the regime of $\gamma\alt 10$~\cite{Tang:2018dq}.  We realize such an array of quasi-1D tube-like traps in a 2D optical lattice by retroreflecting a pair of linearly polarized laser beams.  The lasers are red-detuned by $6.2$~GHz from the $741$-nm Dy narrow-line transition~\cite{Lu:2011gc}. The $\hat{y}$ and $\hat{z}$ lattice beams have waist radii of $150$~$\mu$m and $195$~$\mu$m, respectively. The Dy AC light shift is anisotropic with respect to the magnetic dipole polarization (constrained in the $\hat{x}$--$\hat{z}$ plane) as a result of its large tensor polarizability~\cite{Kao2017}. Therefore, as in previous work~\cite{Tang:2018dq}, we use a half-wave plate to set the polarization of the $\hat{y}$ lattice beam to always be orthogonal to the dipoles to maximize the lattice depth at each $\theta$. By tuning the laser power, we set the lattice depth $V_0 = 30\,E_\text{R}$, where $E_\text{R}/\hbar = 2\pi\times 2.24$~kHz is the recoil energy of a lattice photon. This depth corresponds to a transverse trap frequency of $2\pi\times 25$~kHz and a harmonic oscillator length of $a_\perp=\sqrt{\hbar/m\omega_\perp}=952\,a_0$, where $a_0$ is the Bohr radius.  The typical axial frequency of the tubes is 40~Hz.  The lattice depth is measured using the Kapitza-Dirac diffraction technique~\cite{Gould1986}. The tunneling rate between the tubes is estimated by using 
\begin{equation}
    \frac{J}{E_\text{R}} \simeq \frac{4}{\sqrt{\pi}}\,s^{3/4} \exp(-2\,s^{1/2}),
\end{equation}
where $s = V_0/E_\text{R}$~\cite{Bloch2008}. For our lattice, $s = 30$ and $J/\hbar = 2\pi\times 1$~Hz. Thus, tunneling is negligible during the 100~ms of total time needed to measure collective oscillations.

\subsection{Magnetic field determination} 
\begin{figure}[b!]
    \centering
    \includegraphics[width=\textwidth]{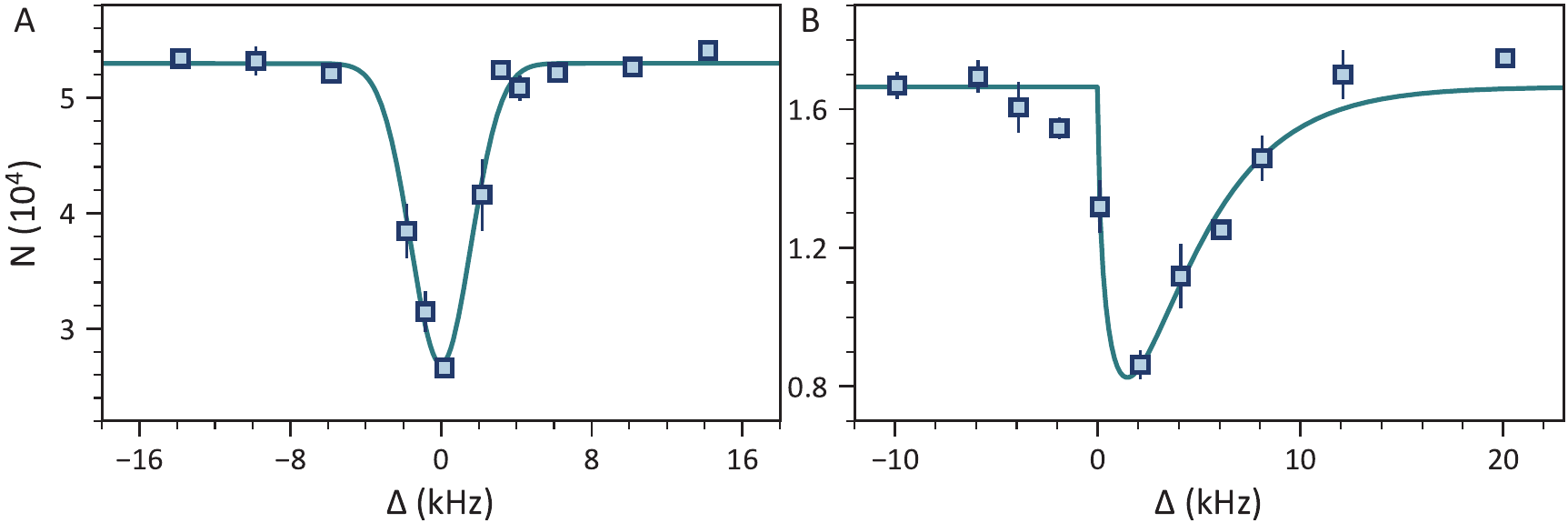}
    \caption{Typical atom loss spectra for (A) magnetic field and (B) molecular binding energy measurements. Here, $N$ and $\Delta$ denote the atom number and RF detuning from the resonant condition, respectively. The solid lines are the best-fit line shape given by a Gaussian and $W_0$ (from Eq.~\eqref{eq:binding_fit}) in the respective panels.}
    \label{fig:spectra}
\end{figure}

To measure the bias field magnitude, we drive  transitions between magnetic sublevels using a weak radiofrequency (RF) field along $\hat{z}$. As a result of dipole-induced spin relaxation~\cite{Burdick2015}, the atoms subsequently release their Zeeman energy as kinetic energy and are ejected from the trap. Resonance appears as an atom loss feature, with the RF frequency equal to the Zeeman energy spacing and, by extension, proportional to the bias field magnitude.  As shown in Supp.~Fig.~\ref{fig:spectra}A, the typical line shape is well approximated by a Gaussian. This procedure is performed after the atoms are loaded into the 2D lattice,  yielding a field accuracy of $\sim$2~mG for all fields and $\theta$-values under consideration.

\subsection{Molecular binding energy spectroscopy} 
The molecular binding energy $E_\text{B}$ in both 3D and quasi-1D can be determined by inducing resonant molecular association~\cite{Jones1999,Thompson2005,Lange2009,Lucioni2018}. Similar to the bias field calibration measurements, we produce an oscillating magnetic field along $\hat{z}$ of the form $B(t)=B_\text{avg}+B_\text{mod}\,\sin(2\pi f_\text{mod} t)$, where $B_\text{mod}$ and $f_\text{mod}$ are the amplitude and frequency of the modulation. The resonant condition ($f_\text{mod}=f_0$) is
%\begin{equation}
    $E_\text{B} + h f_\text{mod} - p_\text{res}^2/m = 0$,
%\end{equation}
where $E_\text{B}$ is the molecular binding energy and $p_\text{res}^2/m$ is the resonant continuum energy. Assuming the Wigner threshold law, the line shape of such a measurement is given by 
\begin{equation}
    W_l(f,f_0) \propto \int_0^\infty \exp\left(\frac{-\epsilon}{T^{'}}\right)\epsilon^{l+1/2}L_\gamma(f,f_0+\epsilon) d\,\epsilon,
\end{equation}
where $\epsilon$ is the collision energy, $L_\gamma(f,f_0+\epsilon)$ is a Lorentzian of full width at half maximum of $\gamma$ centered at $f_0+\epsilon$, and $T^{'}$ is the temperature in units of frequency~\cite{Jones1999}. Assuming only $s$-wave collisions ($l = 0$) and very narrow intrinsic linewidth such that $L_\gamma(f,f_0+\epsilon) \rightarrow \delta\left[\epsilon-(f-f_0)\right]$, we may use the simpler expression
\begin{equation}\label{eq:binding_fit}
    W_0(f,f_0) \propto \exp\left[ -(f-f_0)/T^{'} \right] \left( f-f_0 \right)^{1/2}
\end{equation}
to fit the spectra to the blue side of the resonance. The typical asymmetric line shape is shown in Supp.~Fig.~\ref{fig:spectra}B.

\subsection{Characterization of contact interaction}
Supplemental Fig.~\ref{fig:fr_loss} shows our data revealing the three  Feshbach resonances (FR~1, FR~2, and FR~3) of $^{162}$Dy used in this experiment.  A broader resonance (FR~4) at lower fields (21.93~G) also contributes to the scattering length determination. We conduct high-resolution atom loss spectroscopy across the experimentally relevant field range using an ultracold thermal gas in 3D. Atom loss features at the FR poles are also present under 1D confinement, as shown in Supp.~Fig.~\ref{fig:fr_loss}. We note that FR~1 and FR~4 were first reported in Ref.~\cite{Lucioni2018}.  In particular, the resonance poles of interest for the stiffness and energy measurements are located at $B_{01} \approx 26.90$~G and $B_{02} \approx 27.05$~G, and CIR~1 and CIR~2 are located just to the low-field side of these FRs, respectively.  The coupling between these collisional channels can be neglected since the resonance widths are small, i.e., $\{\Delta_2$, $\Delta_3\}\sim 10$~mG $\ll \Delta_1 \approx 0.2$~G $\ll \Delta_4 \approx 3$~G.  Thus, the magnetic field dependence of the 3D scattering length can be modeled as
\begin{equation}\label{eq:feshbach}
    a_\text{3D}(B) = a_\text{bg} \left(1-\frac{\Delta_1}{B-B_{01}}-\frac{\Delta_2}{B-B_{02}}-\frac{\Delta_3}{B-B_{03}}-\frac{\Delta_4}{B-B_{04}}\right),
\end{equation}
where the background scattering length is denoted by $a_\text{bg}$~\cite{Krzysztof2013}. 

\begin{figure}[t!]
    \centering
    \includegraphics[width=\textwidth]{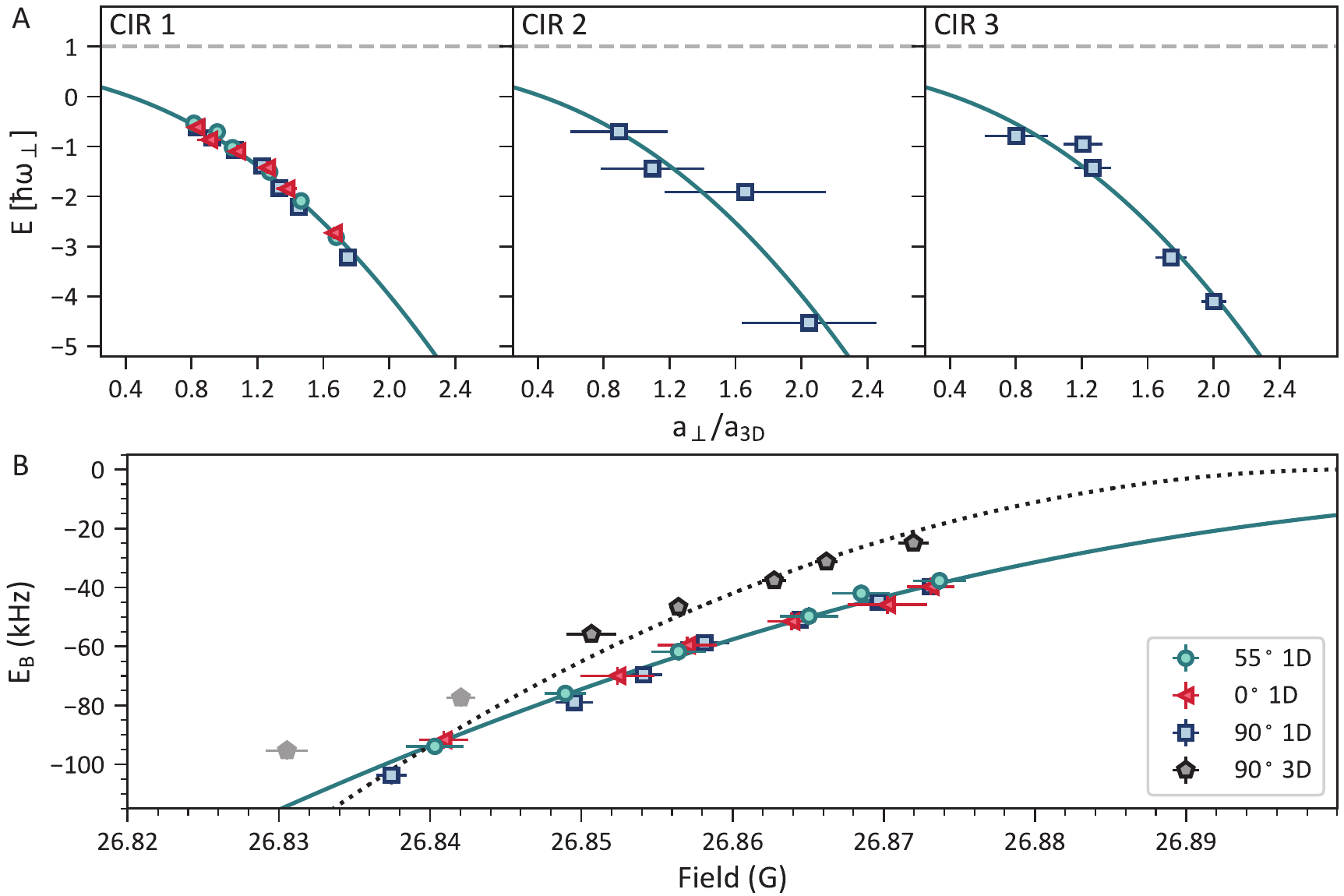}
    \caption{(A) Bound state energy data $E$ under 1D confinement near the three CIRs as a function of the ratio of the transverse harmonic length ($a_\perp$) and 3D scattering length ($a_\text{3D}$).  The open-channel energy is labeled in dashes. (B) 3D and 1D binding energy data $E_\text{B}$ (defined as the energy difference between the open channel and the molecular bound state) around CIR~1 as a function of $B$-field. The best fit to the 3D model using Eq.~\eqref{eq:3D_model} and the 1D model using Eq.~\eqref{eq:1D_model} are shown in dotted and solid lines, respectively. We note that for the 3D data (pentagons), the points without dark edges are outside of the universal regime where $a_\text{3D} \gg \bar{a}$ and  are therefore excluded from the fit.}
    \label{fig:binding_energy}
\end{figure}

Reference~\cite{Lucioni2018} reports two sets of Feshbach parameters based either on anisotropic expansion (AR) data or a combination of AR and $E_\text{B}$ measurements in 3D, as summarized in Table~\ref{tab:feshbach_fit}. We repeat these measurements because, unfortunately, the fit covariance matrix required for error propagation of $a_\text{3D}$ in Eq.~\eqref{eq:feshbach} is not provided in Ref.~\cite{Lucioni2018}.   We conduct our measurements primarily around FR~1 to avoid the multitude of narrow resonance features overlapped with FR~4.  These are too narrow to affect the results of this work, but can complicate the scattering length measurements. Accurate determination of $a_\text{3D}$ depends predominantly on the knowledge of the poles and widths of the three resonances FR 1--3 shown in Supp.~Fig.~\ref{fig:fr_loss}. We note that of the two sets of fit parameters provided in  Ref.~\cite{Lucioni2018}, the AR-only measurements, valid for both $a_\text{3D} > 0$ and $a_\text{3D} < 0$, yield an $a_\text{bg}$ that is within the uncertainty of our previously measured value taken around $5$~G~\cite{Tang2015b,Tang2016}. We choose to use this value. To limit the number of free parameters, we fix the values of $a_\text{bg}$, $B_{04}$, and $\Delta_4$ using this AR-only data. We then extract $B_{0i}$ and $\Delta_i$  and their errors for $i = 1, 2, 3$ from our independent, high-resolution  measurements of $E_\text{B}$  in both 3D and quasi-1D. Similar to Ref.~\cite{Lucioni2018}, we fit the 3D data in the field range $B > 26.85$~G using the corrected universal model
\begin{equation}\label{eq:3D_model}
    E_\text{B}(B) = \frac{\hbar^2}{m\left[a_\text{3D}(B) - \bar{a}\right]^2},
\end{equation}
where  for Dy, the mean scattering length $\bar{a} = 0.956\,R_\text{vdW}$~\cite{Gribakin1993} and $R_\text{vdW} = 80\,a_0$~\cite{Li2016}. For the 1D data, $E_\text{B}$ of the confinement-induced dimers is implicitly given by
\begin{equation}\label{eq:1D_model}
    \frac{a_\text{3D}}{a_\perp} = -\frac{\sqrt{2}}{\zeta(1/2,-E_\text{B}/2\hbar\omega_\perp)},
\end{equation}
where $\zeta$ is the Hurwitz zeta function~\cite{Bergeman2003}. This combined least-square fit yields the six Feshbach parameters and the associated symmetric covariance matrix.  This matrix is given by
\begin{equation}\label{eq:pcov}
    \mathbf{\Sigma} = \kbordermatrix{
& B_{01} & \Delta_1 & B_{02} & \Delta_2 & B_{03} & \Delta_3 \\
B_{01} & 2.2 & 7.2 & 0.1 & 1.2 & -0.1 & -0.4\\ 
\Delta_1 & & 25.2 & 0.2 & 3.4 & -0.0 & -1.2\\ 
B_{02} & &  & 0.3 & 1.0 & -0.0 & -0.0\\ 
\Delta_2 & &  &  & 3.9 & 0.0 & -0.2\\ 
B_{03} & &  &  &  & 0.3 & 0.6\\ 
\Delta_3 & &  &  &  &  & 1.6
},
\end{equation}
in units of mG$^2$. The fit result is shown in Table~\ref{tab:feshbach_fit}, and the $E_\text{B}$ data are summarized in Supp.~Figs.~\ref{fig:binding_energy}.

\begin{table}[t!]
\centering
\begin{tabular}{llllllllll}
\hline
                                                                    & $a_\text{bg}$ ($a_0$) & $B_{01}$ (G) & $\Delta_1$ (G) & $B_{02}$ (G) & $\Delta_2$ (G) & $B_{03}$ (G) & $\Delta_3$ (G) & $B_{04}$ (G) & $\Delta_4$ (G) \\
\hline
Ref.~\cite{Lucioni2018}: AR                   & 180(50)       & 26.892(7)    & 0.14(2)        &              &                &              &                & 21.93(20)    & 2.9(10)        \\
Ref.~\cite{Lucioni2018}: AR + 3D $E_\text{B}$ & 220(50)       & 26.902(4)    & 0.14(5)        &              &                &              &                & 21.91(5)     & 1.9(7)         \\
This work                                                           & 180           & 26.901       & 0.163          & 27.050       & 0.040          & 26.624       & 0.029          & 21.93        & 2.9            \\           
\hline
\end{tabular}
\caption{Summary of Feshbach parameters for the four FRs included in Eq.~\eqref{eq:feshbach}. We choose to report a covariance matrix for the errors in our measurements rather than an error on each entry in the table.   The $6\times 6$ covariance matrix $\mathbf{\Sigma}$ for $B_{0i}$ and $\Delta_i$ for $i = 1, 2, 3$ is given in the text.} \label{tab:feshbach_fit}
\end{table}

\subsection{Excitation of breathing modes} 

\begin{figure}[t!]
    \centering
    \includegraphics[width=\textwidth]{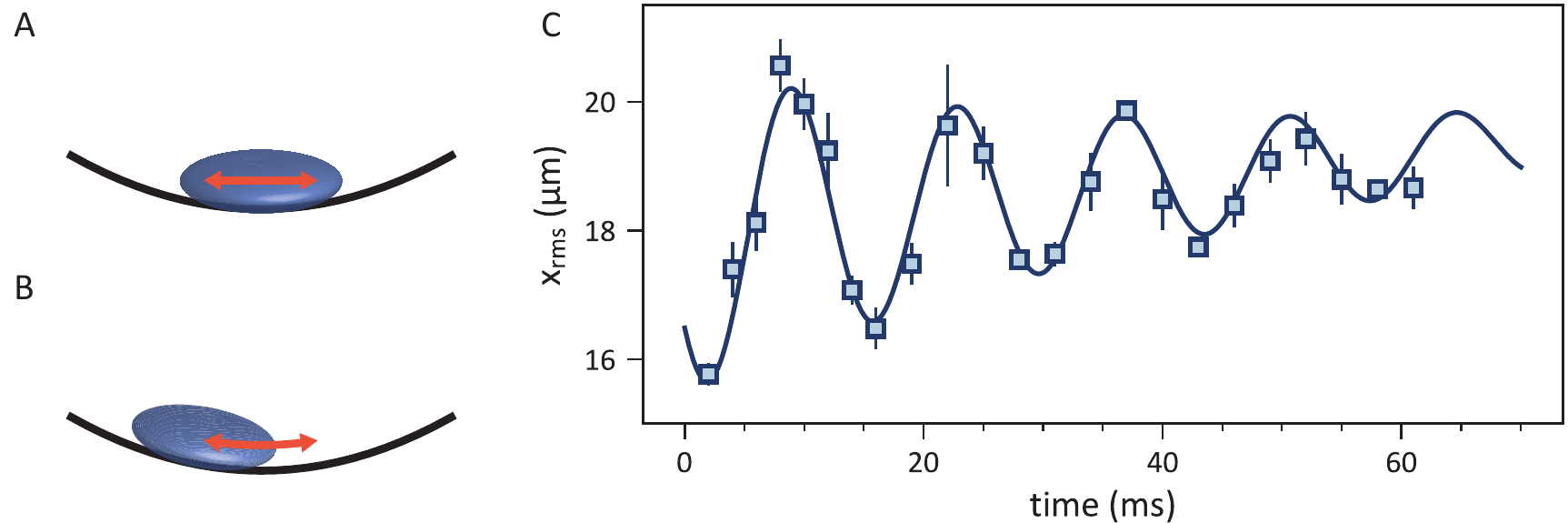}
    \caption{In stiffness measurements, we excite and compare the (A) breathing and (B) dipole modes of the gas confined within a weak harmonic trap along the axial direction $\hat{x}$.  (C) Typical breathing data shows the time evolution of the rms cloud width $x_\text{rms}$ after 14~ms of time-of-flight expansion. The fit to Eq.~\eqref{eq:damped_sine} is shown as a solid line.}
    \label{fig:breathing_example}
\end{figure}

The breathing mode of the trapped quasi-1D gas appears as modulations of the axial width; see Fig.~\ref{fig:breathing_example} for example. The mode may be stimulated by an input perturbation, and there are two such perturbations we employ in this work depending on which is more effective at a particular $g_\text{1D}$; a similar strategy was employed in Ref.~\cite{Haller:2009jrb}. The first method involves quenching the axial trap depth. To do so, we change the trap depth adiabatically with either an additional dipole trap beam or the lattice beams before quickly ramping the laser power back to the pre-quench value.  This induces a breathing of the trapped gas. The second method relies on the fact that the mode can be weakly excited when we perform the $B$-field sweep to tune $a_\text{3D}$ from that used for forced evaporation to its target value. This ramp is performed as fast as possible from the Thomas-Fermi (TF) gas regime, across the TG-regime, over the CIR,  and into the sTG regime, in a similar fashion to Ref.~\cite{Haller:2009jrb}; a fast ramp is desirable due to the need to limit heating from the resonances traversed.  In particular, we set the ramp duration to be $100$~$\mu$s. Once at the target field, we wait $5$~ms for the chamber eddy currents to settle before starting the oscillation measurements. The sweep is sufficiently adiabatic to stimulate no more than a weak excitation of the breathing oscillation and does not cause the system to jump the extensive energy gap between Lieb-Liniger eigenstates shown in Fig.~3.   Moreover, the benignness of the ramp is manifest in the fact that we measure $E/N$'s and $R$'s consistent with that expected from ground states of the repulsive Lieb-Liniger (LL) model (i.e., before the first CIR); see the lowest-energy black circle points in Fig.~3 and Supp.~Fig.~\ref{fig:ground_repulsive}. Sweeping across the discontinuity in $g_\text{1D}$ is nonadiabatic, but continuous in the LL eigenspectrum and wavefunction~\cite{Astrakharchik:2008jz,Chen:2010dl}, which makes possible the smooth transitions from the TF to TG and TG to sTG regimes. The breathing of the gas is studied in time-of-flight using absorption imaging, and we ensure that the resulting breathing amplitude is within 10--20\% of the equilibrium gas width after expansion; see also Ref.~\cite{Haller:2009jrb} for other examples of such measurements in 1D gas systems. In our system, the dipole mode frequency varies between 33 and 47~Hz, whereas the breathing mode frequency lies between 66 and 94~Hz.

\subsection{Fitting of breathing oscillations} 
The expanded gas shape, after integration over the lattice directions, can be approximated by a Gaussian. The time evolution of the best-fit root-mean-square (rms) width can be modeled by a damped sinusoid
\begin{equation}\label{eq:damped_sine}
    x_\text{rms}(t) = a_0\,\exp\left(-\frac{t}{\tau}\right)\sin\left(\omega_\text{B}\,t + \phi\right) + a_1\,t + a_2,
\end{equation}
where $a_0$ is the breathing amplitude, $\tau$ is the damping time of the oscillation, $\omega_\text{B}$ is the breathing frequency, $\phi$ is the breathing phase, $a_1$ encapsulates the increase in gas width due to heating processes, and $a_2$ denotes the equilibrium gas width. We use a standard least-square fitting routine to extract $\omega_\text{B}$. Additionally, to further evaluate the quality of the frequency estimation, we employ a resampling fitting method that is similar to that in Ref.~\cite{Gorg2019}. Specifically, we randomly sample 80\% of the oscillation time series 1000 times. Each of the generated time series is then fit to Eq.~\eqref{eq:damped_sine}. We ensure that the resulting best-fit $\omega_\text{B}$ distribution is single-mode, and that the least-square fitting result lies within $1\sigma$ of the mean of the distribution.
% p[3]*scipy.sin(2*pi*p[1]*x+p[2])*scipy.exp(-x/p[0])+p[4]*x+p[5]

\subsection{Energy per particle measurements} 

The energy per particle $E/N$ is manifest in the width of the momentum distribution. After holding the atoms in the lattice for varying time intervals, we rapidly tune $a_\text{3D}$ to near zero before deloading the lattice using a 500-$\mu$s exponential ramp to release the gas for time-of-flight imaging. This protocol ensures that the expansion of the gas is ballistic. We note that the lattice deloading sequence constitutes a standard band-mapping operation along $\hat{y}$ and $\hat{z}$~\cite{Greiner2001}, which leaves the momentum distribution along the tube axis $\hat{x}$ unaffected. We map-out the time evolution of the axial momentum distribution of the gas when determining the mean width of the expanded gas for use in this  $E/N$ measurement.  This is because state preparation may weakly excite the breathing mode.  We can account for this time-dependent width by measuring  the axial momentum distribution for a varying hold time in the lattice, not just at one hold time.  Only 30~ms is needed to record this time evolution, and a constant-amplitude sinusoid with constant background is used to fit the evolution. This procedure reveals the kinetic part of the energy per particle of the system under investigation, denoted as $E/N$ in the main text. 

The theoretical value of the dimensionless energy density $e(\gamma)$ and kinetic energy density $e_k(\gamma)$ in the ground state can be found by numerically solving the Bethe ansatz equations; see Ref.~\cite{Chen:2010dl} and citations therein. This formalism has been extended to sTG and higher excited states~\cite{Chen:2010dl,yonezawa}. We follow this approach to evaluate $e(\gamma)$ and $e_k(\gamma)$ for all of the parameter space we have experimentally explored, producing the curves in Fig.~3 of the main text and Supp.~Fig.~\ref{fig:second_attractive} below. Systematic shifts due to the underlying harmonic trap and  varying atom number among tubes in our system must be accounted for before the measured $E/N$  may be compared with theory. The first systematic is due to the single-tube harmonic trap. The  kinetic energy of such a trapped system is
\begin{equation}
    \mathcal E_{i,j} = \int \frac{\hbar^2 n^2_{i,j}(x)}{2m} e_k\left[\gamma_{i,j}(x)\right] n_{i,j}(x) dx,
\end{equation}
where $n_{i,j}(x)$ is the density profile for the tube index $(i,j)$ estimated using the local density approximation along with knowledge of $e(\gamma)$~\cite{Dunjko:2001hm, Astrakharchik:2005fq} and $\gamma_{i,j}(x) = -2/a_\text{1D} n_{i,j}(x)$. We then consider tubes with varying atom numbers and perform a weighted average based on the probability $P(M)$ of finding $M$ atoms in a tube~\cite{Paredes:2004fp}. The total resulting calibration factor $C$ can be referenced to the energy density in the central tube as follows
\begin{equation}
    \frac{\mathcal E}{N} = \frac{\hbar^2 n_{0,0}^2}{2m} e_k(\gamma_{0,0}) \cdot C.
\end{equation}
To enable a direct comparison to the theory curve $e_k(\gamma)$, the measured data $E/N$ in Fig.~3 of the main text are normalized by a factor $C\cdot \hbar^2 n_{0,0}^2/2m$ that is recalculated for each set of experimental parameters.

\subsection{Notes on error analysis}

We present some details concerning error analysis. The uncertainties in the dipole and breathing frequencies and $E/N$ values are estimated from least square fit results.   Specifically, these are the diagonal entries of the covariance matrix scaled by the reduced $\chi^2$~\cite{Bevington}. The dominant source of error for the interaction related quantities is $a_\text{3D}$, and the error is propagated using the covariance matrix Eq.~\eqref{eq:pcov} in order to account for correlations between Feshbach parameters. We note that for those data near the zero crossing and point of divergence in $a_\text{3D}$ (e.g., near the limits where $A^2 \rightarrow \infty$ and $0$, respectively), standard error propagation based on a Taylor series  yields a diverging uncertainty. For these cases, we instead use the max-min method that gives asymmetric confidence intervals. This method has  similarly been used in Ref.~\cite{Haller:2009jrb}. Standard error propagation is employed elsewhere~\cite{Bevington,Taylor}.

\section{Supplementary text}

\subsection{Discussion of intertube DDI}

The intertube spacing is 371~nm, half the lattice wavelength. Atoms in nearby tubes are able to interact via the long-range dipole-dipole interaction (DDI)~\cite{Tang:2018dq}. However, these interactions are weak  compared to the total kinetic and short-range interaction energies per atom (outside the weakly repulsive  ground-state regime). Indeed, the collective oscillation data agree with the isolated single-tube predictions for the ground states of the nondipolar repulsive LL model, as demonstrated by the $R$ data in Supp.~Fig.~\ref{fig:ground_repulsive}. Thus, because the intertube DDI evidently plays little role in the repulsive LL model data \textit{and} because the nondipolar 55$^\circ$ sTG data in Fig.~2A  are not too dissimilar from that of the nondipolar Cs system~\cite{Haller:2009jrb}, we conclude that the intertube DDI  plays little role in the attractive LL model data of Figs.~2--4.  That is, both the absolute and differential shifts of the data due to the intertube DDI seem to be small, and we believe that the dramatic, orders-of-magnitude effect on the sTG stability comes from the intratube DDI.

\subsection{Stiffness data for ground states of the repulsive Lieb-Liniger model versus $\theta$}
\begin{figure}[t!]
    \centering
    \includegraphics[width=0.5\textwidth]{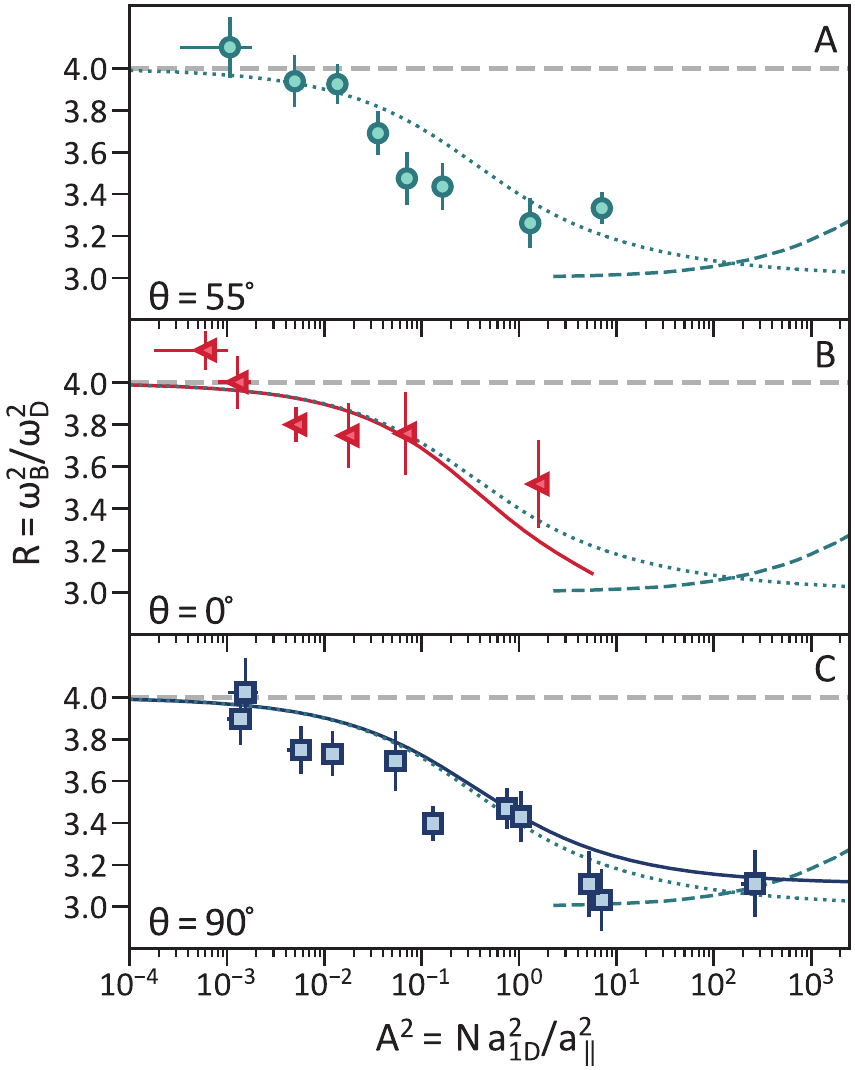}
    \caption{$R$ versus interaction parameter $A^2$ in the repulsive  $g_\text{1D}>0$ regime of the first holonomy cycle. Measurements are shown for the (A) nondipolar ($\theta=55^\circ$), (B) attractive DDI ($0^\circ$), and (C) repulsive DDI ($90^\circ$) systems. Nondipolar theory (dotted line) describing the  TF--to--TG crossover  seems to agree with the data regardless of $\theta$~\cite{Menotti:2002jc}. Accounting for corrections from the regularized 1D DDI provides a small shift from the nondipolar results, as shown in solid lines in panel (B) and (C). These lines are similar to that calculated in a recent perturbative treatment~\cite{DePalo:2020}.  Regardless, these shifts are too small to be resolved by the experiment. We additionally show results of a nondipolar Hartree calculation (dashes) that describes the crossover of the TF gas into the weakly interacting regime of a Gaussian BEC~\cite{Astrakharchik2015}.}
    \label{fig:ground_repulsive}
\end{figure}

For nondipolar gases, it is known  that $R$ transitions from 3 for a TF BEC ($A^2$ roughly between 1 and 100) before rising to 4 as $A^2\rightarrow 0$ in the TG limit~\cite{Menotti:2002jc}. This expectation is borne out by our $55^\circ$ data in Supp.~Fig.~\ref{fig:ground_repulsive}A and those of the Cs experiment~\cite{Haller:2009jrb}.  Moreover, this seems to be true regardless of $\theta$, showing how little the DDI affects the ground state system.

\subsection{Stiffness data for 90$^\circ$ excited states on the attractive branch of the second holonomy level}
\begin{figure}[t!]
    \centering
    \includegraphics[width=0.5\textwidth]{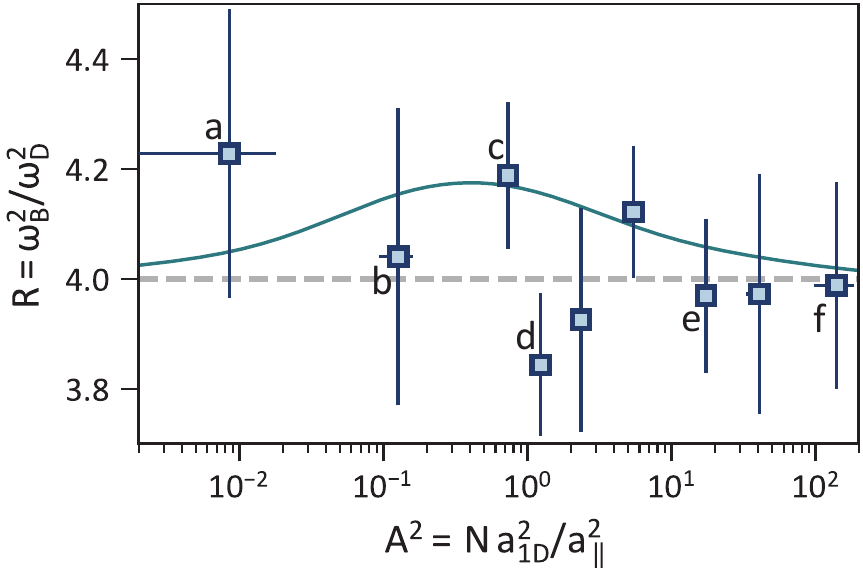}
    \caption{Stiffness $R$ versus interaction parameter $A^2$ in the attractive $g_\text{1D}<0$ regime of the second holonomy cycle for the repulsive DDI ($\theta=90^\circ$) system. Letter labels refer to the points in Figs.~1 and 3 of the main text. The solid curve is the Bethe ansatz prediction.}
    \label{fig:second_attractive}
\end{figure}

We show stiffness data for the attractive branch of the second holonomy level in Supp.~Fig.~\ref{fig:second_attractive}. The Bethe ansatz prediction is also plotted~\cite{Chen:2010dl}. Since thermal gases have the same stiffness, $R=4$ (see inset of Supp.~Fig.~\ref{fig:heating}B)~\cite{Menotti:2002jc,Moritz:2003kv}, collective oscillation measurements cannot alone tell us whether these states are nonthermal. However, we can discern their nonthermal nature by sweeping in and out of this regime for various hold times before making measurements of $E/N$ and $R$, as shown in Fig.~4 of the main text.

\vspace{-3mm}
\subsection{Absence of CIR shift}
Several theoretical works predict a dependence of the CIR position on the DDI strength and/or angle~\cite{Shi:2014ki,Guan:2014cc,Giannakeas:2013fd,Sinha:2007gx}.  We do not observe any such shift in our molecular binding energy measurements, within experimental resolution.  This might be due to the narrowness of the Feshbach resonances in Dy, or to their unusually complicated character~\cite{Petrov:2012bq}, both not considered in those works.

\subsection{Theory predictions of purely dipolar sTG gases}

Theoretical results prior to the start of this work focused exclusively on purely dipolar models, in which the only interaction comes from the DDI~\cite{Arkhipov:2005gg,Citro:2007gs,Astrakharchik:2008jz,Pedri:2008hi,Deuretzbacher:2010gg,Deuretzbacher:2013da,Girardeau:2012be}. As such, these cannot describe the case at hand wherein there is an interplay between the short-range contact interaction and the DDI and it is the cyclical manipulation of the contact that topologically pumps the system.

Contact-free, repulsive DDI sTG gases were predicted to exist in the ground state, if one takes the definition of the sTG to be a gas with correlations stronger than that of a TG gas, regardless of whether it is excited~\cite{Arkhipov:2005gg,Citro:2007gs,Astrakharchik:2008jz,Pedri:2008hi,Deuretzbacher:2010gg,Deuretzbacher:2013da}.  Such correlations arise due to the long-range nature of the repulsive DDI, and not to quenching into any attractive potential.  Such states are very different from the DDI-stabilized \textit{excited}-state quenched  sTG gases discussed here. Regardless, the repulsive DDI strength of Dy is insufficient to induce ground-state sTG gases at the densities explored.  Reference~\cite{Girardeau:2012be} did consider quenching a purely dipolar gas into an excited state sTG gas by abruptly rotating the magnetic field angle from repulsive ($\theta = 90^\circ$) to attractive ($\theta = 0^\circ$). But again, they did not consider the simultaneous effect of a contact interaction or its use for topological pumping. 

Stimulated by our work, the authors of Ref.~\cite{DePalo:2020} considered the repulsive \textit{ground} state of a Lieb-Liniger model with the DDI perturbatively added. They report $R$ versus $A^2$ curves very similar to those presented as solid lines in Supp.~Fig.~\ref{fig:ground_repulsive} that were calculated using a regularized 1D DDI.

\subsection{Regularized 1D dipole-dipole interaction}

References~\cite{Arkhipov:2005gg,Citro:2007gs,Astrakharchik:2008jz,Pedri:2008hi,Girardeau:2012be} employed the unregularized DDI in their 1D models. This contains the $1/r^3$ divergence when two atoms approach one another. However, in a quasi-1D trap, this divergence is smoothed out by the transverse degrees of freedom; the interaction must be regularized by integrating out these transverse degrees of freedom~\cite{Deuretzbacher:2010gg,Deuretzbacher:2013da,Tang:2018dq,DePalo:2020}. This operation removes the divergence, replacing it with an extra delta function-like term while reducing the strength of the $1/r^3$ term at short distances. 

We now address whether there might be any contribution of this effective short-range term to our experiment. Such a contribution would be manifest as a shift of the CIR versus $\theta$. As we do not measure any such shift, nor any inconsistency with respect to nondipolar theory in the ground state, we conclude that the effect is negligible in our mapping of $B$ to $a_\text{3D}$. Thus, $g_\text{1D}$ and $\gamma$ are unaffected. Away from the CIR, the contact-like DDI contribution may be simply added to the van der Waals contribution, as in Ref.~\cite{Tang:2018dq}. We follow this method to: 1) generate the solid $R$ versus $A^2$ curves in Supp.~Fig.~\ref{fig:ground_repulsive}, which are nearly indistinguishable from that derived by a perturbative calculation presented in Ref.~\cite{DePalo:2020}; and 2) determine the position of the vertical dotted line in Fig.~2C of the main text and in Supp.~Fig.~\ref{fig:minimal_state}.  An axial trap frequency of 40~Hz and $N = 35$ are used to determine the position.

\subsection{Minimal state}

\begin{figure}[t!]
    \centering
    \includegraphics[width=0.5\textwidth]{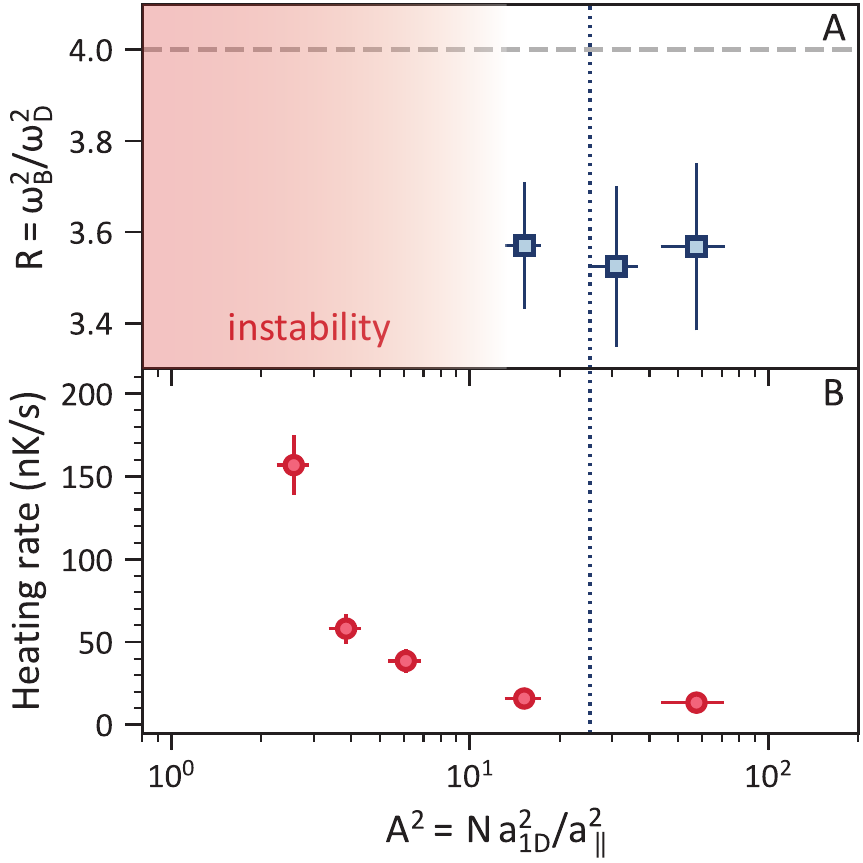}\vspace{-3mm}
    \caption{A minimal state~\cite{yonezawa} can be demonstrated by inverse cycling $g_\text{1D}$ from positive to negative through zero (and not through the CIR-induced divergence), as described in the text. Stiffness and heating rate measurements for $g_\text{1D} < 0$ are shown in panel (A) and (B), respectively. The vertical dotted line shows the point where the short-range DDI and van der Waals contributions to the 1D coupling strength are equal and opposite. The system becomes susceptible to collapse beyond this point (i.e., toward smaller $A^2$), where the attractive contact interactions begin to dominate the repulsive DDI.  This collapse is manifest  as a diverging heating rate and an $R$ that softens to 0.  In panel (A), breathing mode oscillations cannot be observed for $A^2$ less than $\sim$10 due to this collapse instability.}
    \label{fig:minimal_state}
\end{figure}

In the main text, we considered cycles in which (in the ideal limit) each eigenstate is pumped to a higher-energy many-body eigenstate. We can also consider the reverse cycle: under this, some eigenstates map to ``collapsed'' states with large numbers of molecular clusters and a corresponding divergence of the energy in the unitary limit. These minimal states are defined as  eigenstates of the Hamiltonian for which this inverse cycle leads to a divergence of eigenenergy~\cite{yonezawa}. In other words, this is a ground state that is tuned into a bound state via cycling $g_\text{1D}$ backwards:  tuning $g_\text{1D} > 0$ to $g_\text{1D} <0$ directly (i.e., without crossing a CIR) induces a collapse just like in attractive BECs in higher dimensions. We demonstrate that in our system, the states with $g_\text{1D} \rightarrow 0^+$ in the first holonomy level are  minimal states by tuning them directly to $g_\text{1D} \rightarrow 0^-$ by crossing  $g_\text{1D} = 0$. To do so, we produce a 1D BEC at a $B$-field of 29.805~G (beyond CIR~2) for the repulsive DDI configuration (90$^\circ$) before  performing the inverse cycle of $g_\text{1D}$ from positive to negative through zero. This protocol ensures that the gas does not get pumped to higher holonomy levels via crossing a CIR and the resulting $g_{1D}\rightarrow \pm \infty$ divergence. As shown in Supp.~Fig.~\ref{fig:minimal_state}B, the heating rate diverges shortly after the point where the van der Waals (attractive) and DDI (positive) contributions to $g_\text{1D}$ cancel, indicating the onset of collapse due to  bound-state formation. This is confirmed by the stiffness measurement in Supp.~Fig.~\ref{fig:minimal_state}A, where we are unable to excite a well-defined breathing mode below the same $A^2$ at which the heating begins to significantly increase.

\subsection{Heating measurements} \vspace{-3mm}

\begin{figure}[t!]
    \centering
    \includegraphics[width=\textwidth]{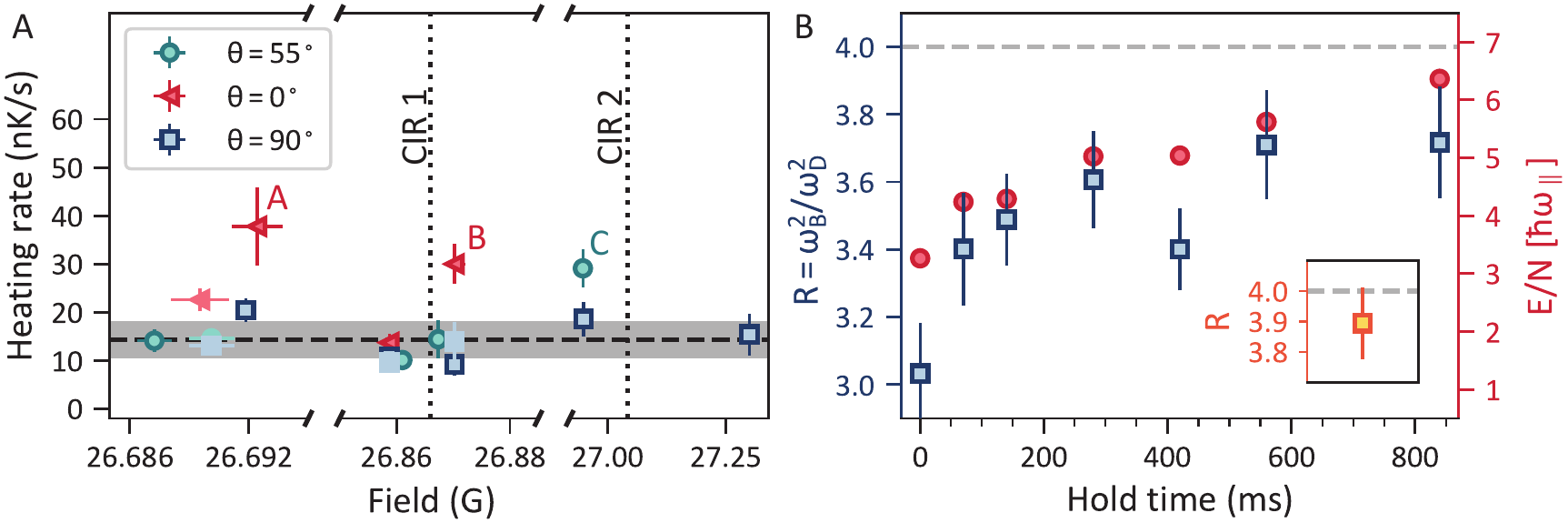}
    \caption{(A) Heating rate in 1D is plotted as a function of magnetic field, which provides tunability across different interaction regimes. The circles, triangles, and squares denote $\theta = 0^\circ$, $55^\circ$, and $90^\circ$, respectively. The points with dark edges are measured after stimulating the breathing mode, whereas those without are at rest. The mean heating rate excluding the labeled outliners is shown in dashes, and the shaded area spans $\pm 1\sigma$ (standard deviation). Details regarding the three outlier points in the unstable regime for $\theta = 0^\circ$ and $55^\circ$ (labeled A, B, and C) are provided in the text. (B) Heating of a TF gas with $\theta = 90^\circ$ and $A^2 = N\,a_\text{1D}^2/a_\parallel^2 \approx 7$. The ratio $R =(\omega_\text{B}/\omega_\text{D})^2$ (squares) and kinetic energy per particle $E/N$ (circles) are plotted as a function of the hold time before breathing oscillations are stimulated and measured.  These data may be compared to the thermalization (without heating) of excited states  in Fig.~4 of the main text. The inset shows the stiffness of a thermal gas at 160 nK.}
    \label{fig:heating}
\end{figure}

We determine the heating rate of the system for all interaction strengths in the ground and excited states.  These measurements are summarized in Supp.~Fig.~\ref{fig:heating}A. The experimental sequence is identical to the breathing mode measurements, and the kinetic energy per particle $E/N$ for each hold time can be extracted from the squared RMS width of the expanded gas, whose profile along the tube direction is well approximated by a single Gaussian. 

Heating seems to ensue only after a hold time of $t_0 \approx 100$~ms, which is longer than the final time of the collective oscillation measurements. The fitted width remains roughly constant before $t_0$.  The $E/N$ increases linearly after this, and the best-fit slope determines the heating rate.  We do not observe enhanced heating when the gas is quenched into a breathing mode. Both observations suggest that thermal effects are of  negligible influence to the breathing mode measurements.

While most measured heating rates are within $1\sigma$ of the mean value of $14$~nK/s, some remarks on the outliers are in order. First, when $\theta = 0^\circ$, the dipoles are in the prolate configuration (i.e., they are aligned with the symmetry axis of the cylindrical tubes). The DDI is attractive in such a trap where the trap aspect ratio $\lambda = \omega_x/\omega_\perp < 1$, leading to an instability at the mean-field level of the dipolar BEC, even in the presence of a repulsive contact interaction. (See Ref.~\cite{Edler:2017bg} for a discussion of fluctuation corrections.)  In our system, $\lambda = 0.002 \approx 0$, and the critical 3D scattering length $a_\text{3D, crit}$ is approximately equal to the dipolar length $a_\text{dd} = C_\text{dd}\,m/12\pi\hbar^2 = 129\,a_0$, where $C_\text{dd} = \mu_0\mu^2 = \mu_0 (9.93\mu_\text{B})^2$ is the DDI coupling constant for Dy, $\mu_0$ is the vacuum permeability, and $\mu_\text{B}$ is Bohr magneton~\cite{Lahaye2009}. For the point labeled A in the TF regime, $a_\text{3D} = 154(6)\,a_0$ and, the enhanced heating rate when the breathing oscillations are initiated may be attributed to this mean-field mechanical instability. The source of heating for points B and C might  be due to many-body cluster formation, since they are near the collapse point.

\vspace{-3mm}
\subsubsection{Thermally driven crossover from a Thomas-Fermi to an ideal Bose gas} 

We study  thermal effects on the equation-of-state of a TF gas in the ground state as an independent means of verifying the characteristic heating rate of the quasi-1D system. Specifically,  we set $\theta = 90^\circ$ and $A^2 \approx 7$ (before crossing CIR~1), where $R \approx 3$ suggests that the gas is deep within the TF regime. We hold the atoms in the lattice for a varying delay time  before initiating the breathing oscillation measurement. As predicted by Yang-Yang thermodynamics~\cite{Yang1969}, the transition from the TF ($R = 3$) to the ideal Bose gas ($R = 4$) regime occurs at a crossover temperature $T_\text{co} = \sqrt{\gamma}\,\hbar^2 n_\text{1D}^2 / 2m$, where $\gamma = 2/n_\text{1D}a_\text{1D}$ is the  dimensionless LL interaction parameter and $n_\text{1D}$ is the 1D linear density~\cite{Jacqmin2011}. For our experimental parameters, $T_\text{co} \approx 17$~nK.   The data in Supp.~Fig.~\ref{fig:heating}B shows that it takes over 840~ms for  $R$ to rise to 4.  Given the typical heating rate of $14$~nK/s (see Supp.~Fig.~\ref{fig:heating}A),  it would take $\sim$1~s to cross over into the ideal Bose gas regime if we assume the  initial gas temperature is much less than $T_\text{co}$. This timescale is consistent with our data.

\vspace{-3mm}
\subsection{Comment regarding comparison of 55$^\circ$ and nondipolar Cs data in Ref.~\cite{Haller:2009jrb}}

We first restate qualitative arguments regarding where along $A^2$ the collapse should occur for nondipolar gases~\cite{Astrakharchik:2004ht}.  The point roughly occurs when the sum length $Na_\text{1D}$ equals the width of the harmonic container, which is proportional to $\sqrt{N}a_\parallel$ for fermionized atoms; i.e., when $A = \sqrt{N}a_\text{1D}/a_\parallel = 1$. The collapse may be described in analogy to the classical gas of hard rods of length $a_\text{1D}$~\cite{Astrakharchik:2004ht}.  The classical gas becomes infinitely stiff as the sum of the rod lengths $Na_\text{1D}$ equals the length of the container, whereas the  stiffness of the quantum gas softens to zero in the collapse. The earlier collapse of our $0^\circ$ data may be viewed in this picture as arising from the attractive DDI `elongating the rods.'

Our nondipolar data for 55$^\circ$ (see Fig.~2A) is shifted by $\sim$10$\times$ to the low-$A^2$ side with respect to the Cs data  in Ref.~\cite{Haller:2009jrb}, itself shifted somewhat to the low-$A^2$ side of the estimate given above (and the variational Monte Carlo data).  The reason for these shifts is unknown.  Theory is not yet able to  predict where the data should intercept $R=0$ for any trapped quantum system~\cite{Astrakharchik2019}, let alone the dipolar gas, as this has to do with details of the complicated coupling to the multitude of cluster states. Variational Monte Carlo suggests the gas becomes unstable ($R\rightarrow0$) at $A^2=0.6$, but this is only approximate~\cite{Astrakharchik:2005fz}. The shift is likely nonuniversal in origin, as is the  exact intercept of the data with $R=4$~\cite{Astrakharchik2019}.

\end{document}